\documentclass[a4paper,dvips]{jpconf}
\usepackage[dvips]{epsfig}
\usepackage{axodraw_c,color} 
\usepackage{graphics}
\usepackage{wrapfig}
\usepackage{floatfig}

\newcommand{\tfrac}[2]{{\textstyle{\frac{#1}{#2}}}}

\definecolor{green}{rgb}{0,0.7,0}
\definecolor{blue}{rgb}{0,0,1}
\definecolor{red}{rgb}{1,0,0}
\definecolor{brown}{rgb}{0.7,0.3,0}
\definecolor{violet}{rgb}{0.5,0,0.5}

\begin{document}
\title{Four-point function in general kinematics through geometrical splitting and reduction}

\author{Andrei I Davydychev}

\address{Institute for Nuclear Physics, Moscow State University,
119992 Moscow, Russia}

\ead{davyd@theory.sinp.msu.ru}

\begin{abstract}
It is shown how the geometrical splitting of $N$-point Feynman diagrams 
can be used to simplify the parametric integrals and reduce the number 
of variables in the occurring functions. As an example, a calculation 
of the dimensionally-regulated one-loop four-point function in general 
kinematics is presented.
\end{abstract}

\vspace*{-130mm}
\begin{flushright}
MSU-SINP 2017-1/891\\
$\;$\\
\footnotesize{
Contribution to the Proceedings of ACAT-2017\\ 
(Seattle, USA, August 21--25, 2017)
}
\end{flushright}
\vspace*{105mm}

\section{Introduction}

\begin{wrapfigure}{rb}{0.5\textwidth}
\vspace*{-2mm} 
  \begin{center}
  \vspace*{4mm}
  \includegraphics[width=18pc]{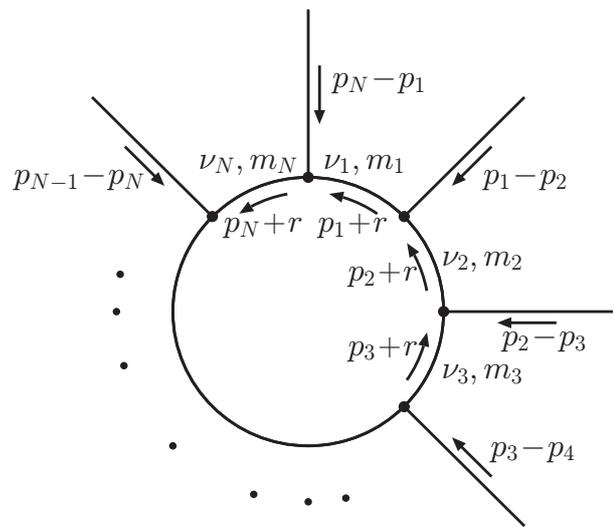}
  \vspace*{5mm}
  \caption{\label{label}$N$-point one-loop diagram}
  \label{figNpt1}
  \vspace*{-8mm} 
  \end{center}
\end{wrapfigure}

In the general off-shell case, one-loop $N$-point 
diagrams (shown in figure~\ref{figNpt1})
depend on $\tfrac{1}{2}N(N-1)$ momentum invariants $k_{jl}^2=(p_j-p_l)^2$
and $N$ masses of the internal particles $m_i$. Here and below, 
for the corresponding scalar integrals we follow
the notation $J^{(N)}\left(n; \{\nu_i\}|\{k_{jl}^2\},\{m_i\}\right)$
used in~\cite{JMP1-2}, where $\nu_i$ are the
powers of the internal scalar propagators, and 
the space-time dimension is denoted as $n$, so that we can also deal
with the dimensionally-regulated integrals with $n=4-2\varepsilon$ \cite{dimreg}.
Below we will mainly consider the cases when all $\nu_i=1$.  

A geometrical interpretation of kinematic invariants and other quantities
related to $N$-point Feynman diagrams helps us 
to understand the analytical structure of the results for these diagrams.
As an example, singularities of the general three-point function can be 
described pictorially through a tetrahedron constructed out of
the external momenta and internal masses. 
Such a geometrical visualization can be used to derive Landau equations defining
the positions of possible singularities \cite{Landau} 
(see also in \cite{3pt_sing}). 

In~\cite{DD-JMP,D-NIMA,Crete} it was demonstrated how such geometrical
ideas could be used for an analytical calculation of
one-loop $N$-point diagrams. For the geometrical interpretation, a ``basic simplex"
in $N$-dimensional Euclidean space is employed (a triangle for $N=2$, 
a tetrahedron for $N=3$, etc.), and the obtained results are expressed
in terms of an integral over an $(N-1)$-dimensional spherical (or hyperbolic) 
simplex, which corresponds to the intersection of the basic simplex and the 
unit hypersphere (or the corresponding hyperbolic hypersurface), 
with a weight function depending on the angular distance $\theta$ between 
the integration point and the point 0, 
corresponding to the height of the basic simplex
(see in~\cite{DD-JMP}). 
For $n=N$ this weight function is equal to 1, and the results simplify: for the case $n=N=3$
see in~\cite{Nickel}, and for the case $n=N=4$ see in~\cite{OW,Wagner}. Other interesting
examples of using the geometrical approach can be found, e.g., in~\cite{other_geom}.

In this paper we will demonstrate that the natural way of splitting the basic simplex, 
as prescribed within the geometrical approach discussed above, leads to a reduction
of the effective number of independent variables in separate contributions obtained 
as a result of such splitting~\cite{D-JPCS}. Moreover, by considering examples 
with $N\leq4$ we will show that this reduction leads to simplifications in the
corresponding Feynman parametric integrals, which can be explicitly calculated 
in terms of the (generalized) hypergeometric functions. 

\section{Two-point function}

\begin{figure}[t]
\begin{center}
\vspace*{-5mm}
\includegraphics[width=30pc]{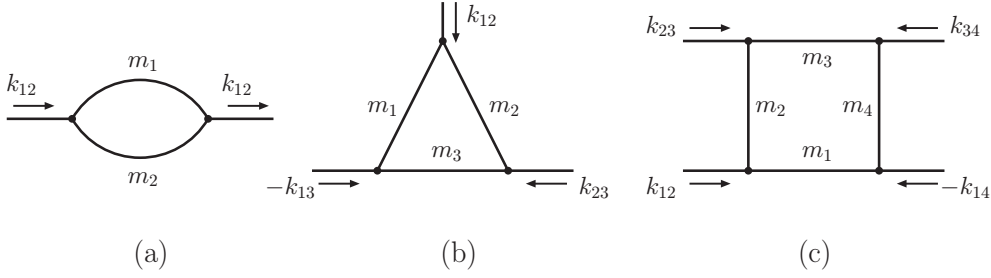}
\end{center}
\vspace*{-5mm}
\caption{\label{label}Momenta and masses in (a) two-point (b) three-point and (c) four-point diagrams.}
\label{fig234pt}
\end{figure}

\begin{wrapfigure}{rb}{0.5\textwidth}
\vspace*{1mm}
\begin{center}
\includegraphics[width=16pc]{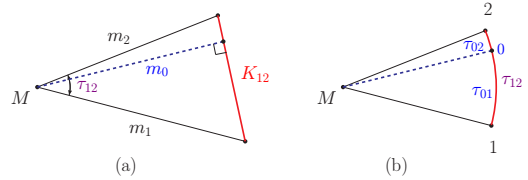}
\vspace*{-5mm}
\caption{\label{label}Two-point case: (a) the basic triangle and (b) the arc $\tau_{12}$.}
  \label{fig2pt1}
\vspace*{-8mm}
\end{center}
\end{wrapfigure}

%

For the two-point function (see figure~\ref{fig234pt}a), 
there is only one external momentum invariant $k_{12}^2$,
and the sides of the corresponding basic triangle are $m_1$, $m_2$ and $K_{12}\equiv\sqrt{k_{12}^2}$
(see figure~\ref{fig2pt1}a). The angle $\tau_{12}$ between the sides $m_1$ and $m_2$
is defined through $\cos\tau_{12}\equiv c_{12}=(m_1^2+m_2^2-k_{12}^2)/(2m_1 m_2)$, and 
(in the spherical case) the integration goes over the arc $\tau_{12}$ of the unit circle,
as shown in figure~\ref{fig2pt1}b.

For splitting we use the height of the basic triangle, 
$m_0=m_1 m_2 \sin{\tau_{12}}/\sqrt{k_{12}^2}$, and obtain two triangles with 
the sides ($m_1$, $m_0$, $K_{01}\equiv\sqrt{k_{01}^2}$) and 
($m_2$, $m_0$, $K_{02}\equiv\sqrt{k_{02}^2}$),
respectively. By construction, $K_{01}+K_{02}=K_{12}$. Here
$k_{01}^2 = (k_{12}^2+m_1^2-m_2^2)^2/(4 k_{12}^2)$ and
$k_{02}^2 = (k_{12}^2-m_1^2+m_2^2)^2/(4 k_{12}^2)$ (note that $k_{01}^2=m_1^2-m_0^2$ and
$k_{02}^2=m_2^2-m_0^2$). Each of the resulting integrals
can be associated with a two-point function, and we arrive 
at the following decomposition~\cite{D-JPCS}:  
\begin{eqnarray}
J^{(2)}\left(n; 1,1| k_{12}^2; m_1, m_2 \right) &=&
\frac{1}{2k_{12}^2}
\left\{
(k_{12}^2+m_1^2-m_2^2)
J^{(2)}\left(n; 1,1| k_{01}^2; m_1, m_0 \right)
\right.
\nonumber \\ && \qquad
\left.
+ (k_{12}^2-m_1^2+m_2^2)
J^{(2)}\left(n; 1,1| k_{02}^2; m_2, m_0 \right)
\right\} \; .
\end{eqnarray}
This is an example of a functional relation between integrals
with different momenta and masses, similar to those described in~\cite{Tarasov2}.
Moreover, as shown in \cite{D-JPCS}, we can represent 
the right-hand side in terms of the equal-mass integrals 
$J^{(2)}\left(n; 1,1| 4 k_{01}^2; m_1, m_1 \right)$ and 
$J^{(2)}\left(n; 1,1| 4 k_{02}^2; m_2, m_2 \right)$.

Let us look at the number of variables. In the original integral 
$J^{(2)}\left(n; 1,1| k_{12}^2; m_1, m_2 \right)$ we have three 
independent variables: two masses and one momentum invariant 
(out of them we can construct two dimensionless variables).
In the integral $J^{(2)}\left(n; 1,1| k_{01}^2; m_1, m_0 \right)$
we have one extra condition on the variables, $k_{01}^2=m_1^2-m_0^2$,
so that we get two independent variables (i.e., 
one dimensionless variable). 

This can also be seen in the integrands of Feynman parametric integrals:
for the original two-point integral, the quadratic form is 
\begin{equation}
J^{(2)}\left( n; 1,1\big| k_{12}^2; m_1, m_2 \right)
\Rightarrow
[ \alpha_1\alpha_2 k_{12}^2 - \alpha_1 m_1^2 - \alpha_2 m_2^2 ] \; ,
\end{equation}
whereas for one of the resulting integrals after splitting, 
remembering that $\alpha_1+\alpha_2=1$, we get
\begin{equation}
J^{(2)}\left( n; 1,1 \big| k_{01}^2; m_1, m_0\right) \Rightarrow  
[ \alpha_1\alpha_2 k_{01}^2 - \alpha_1 m_1^2 - \alpha_2 m_0^2 ] 
= - [ \alpha_1^2 k_{01}^2 + m_0^2 ]\; .
\end{equation}
In this way, we obtain the following result in arbitrary dimension:
\begin{eqnarray}
J^{(2)}\left( n; 1,1 \big| k_{01}^2; m_1, m_0\right) &=& 
{\rm i} \pi^{n/2} \Gamma(2-n/2)
\int\limits_0^1 \!\! \int\limits_0^1 
\frac{{\rm d}\alpha_1 \; {\rm d}\alpha_2 \; \delta(\alpha_1+\alpha_2-1)}
{[ \alpha_1^2 k_{01}^2 + m_0^2 ]^{2-n/2}}
\nonumber \\
&=& {\rm i}\pi^{n/2} \frac{\Gamma(2-n/2)}{(m_0^2)^{2-n/2}}\;
{}_2F_1\left(
\begin{array}{c} 1/2,\; 2-n/2\\ 3/2 \end{array} \Biggl| -\frac{k_{01}^2}{m_0^2} 
\right) \; ,
\end{eqnarray}
where $_2F_1$ is the Gauss hypergeometric function.
Similar expression for the second integral, 
$J^{(2)}\left( n; 1,1 \big| k_{02}^2; m_0, m_2\right)$, can be obtained by 
permutation $1\leftrightarrow2$.
Therefore, the result for the two-point function in arbitrary dimension 
can be expressed in terms of a combination of two $_2F_1$ functions of
a single dimensionless variable (see, e.g., in~\cite{DD-JMP,BDS}) 
whose $\varepsilon$-expansion is known to any order~\cite{D-ep,DK1}. 

\section{Three-point function}

\begin{figure}[b]
\begin{minipage}{18pc}
\vspace*{10mm}
\includegraphics[width=18pc]{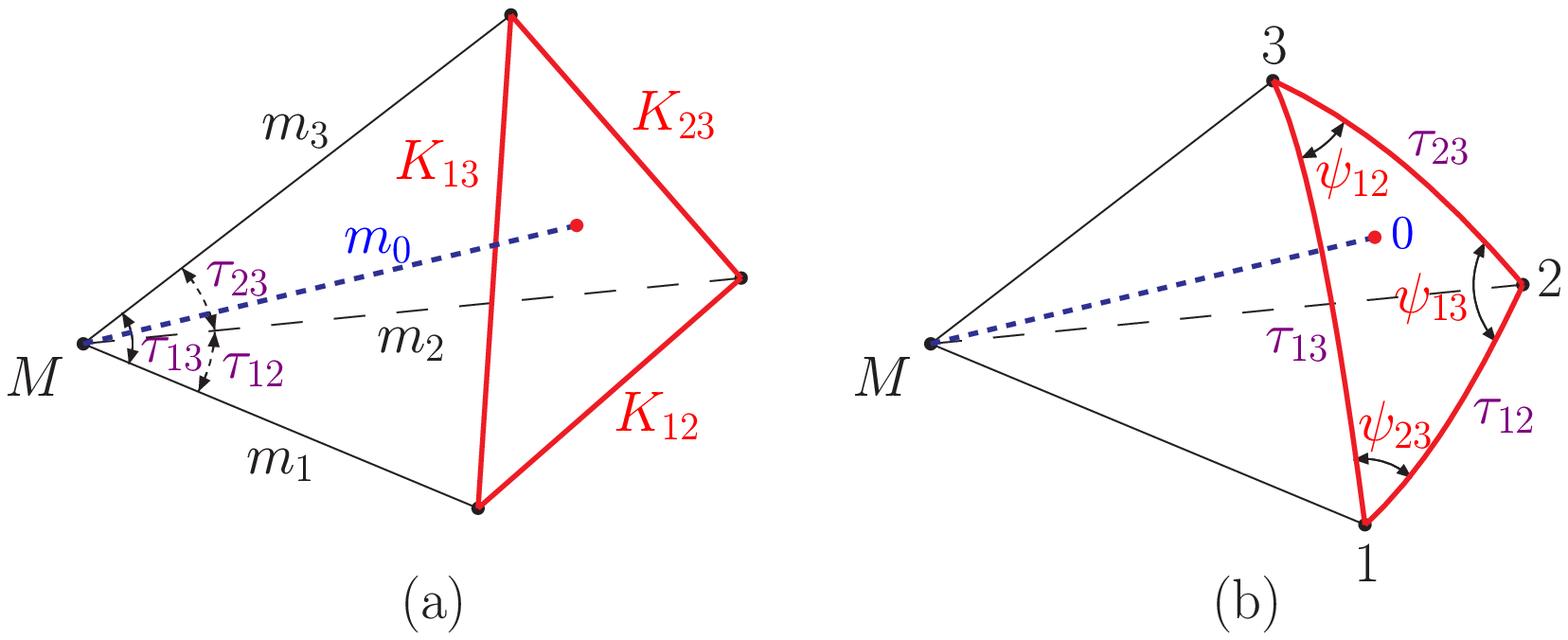}
\vspace*{-15mm}
\caption{\label{label}Three-point case: (a) the basic tetrahedron 
and (b) the solid angle.}
  \label{fig3pt1}
\end{minipage}\hspace{2pc}%
\begin{minipage}{18pc}
\vspace*{11mm}
\includegraphics[width=18pc]{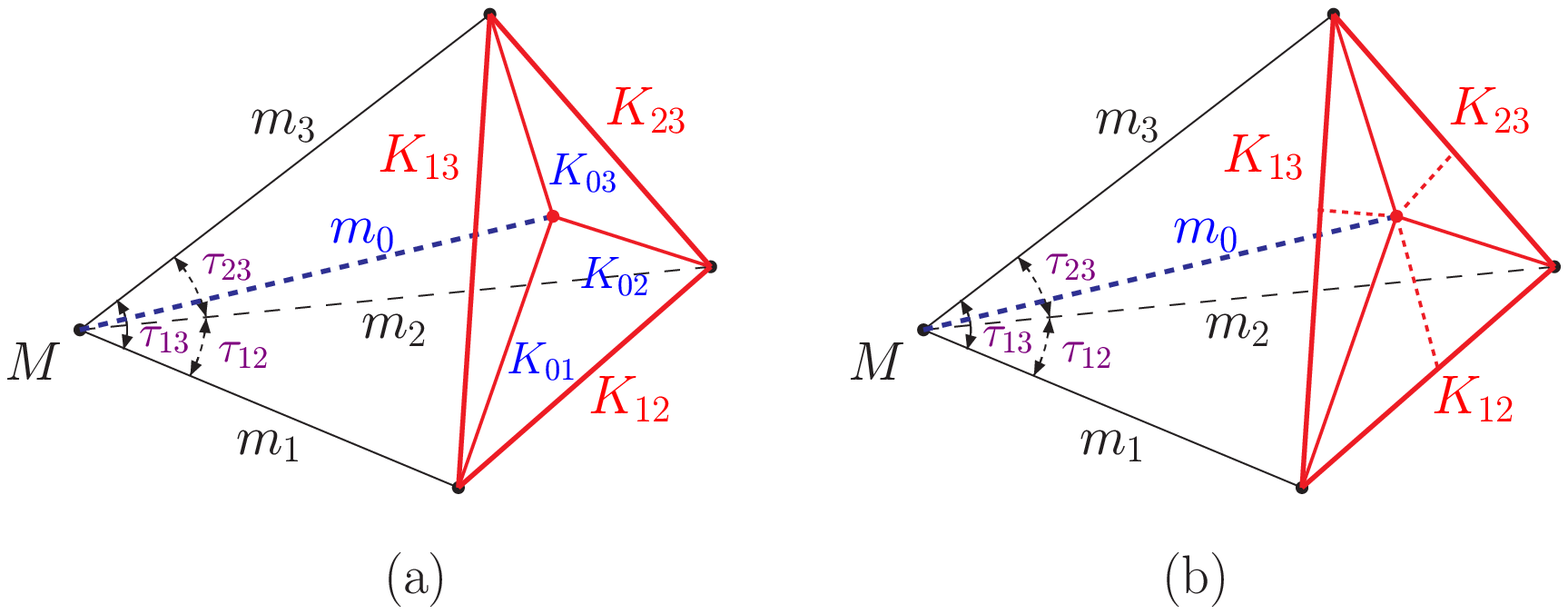}
\vspace*{-15mm}
\caption{\label{label}(a) Splitting the basic tetrahedron into three tetrahedra 
and (b) further splitting into six tetrahedra.}
  \label{fig3pt2}
\end{minipage} 
\end{figure}

For the three-point function (see figure~\ref{fig234pt}b), 
there are three external momentum invariants, $k_{12}^2$,
$k_{13}^2$ and $k_{23}^2$,
and the sides of the corresponding basic tetrahedron are $m_1$, $m_2$, $m_3$, 
$K_{12}\equiv\sqrt{k_{12}^2}$,
$K_{13}\equiv\sqrt{k_{13}^2}$ and $K_{23}\equiv\sqrt{k_{23}^2}$
(see figure~\ref{fig3pt1}a). The angles $\tau_{12}$, $\tau_{13}$ and $\tau_{23}$
between the sides $m_1$, $m_2$ and $m_3$
are defined through $\cos\tau_{jl}\equiv c_{jl}=(m_j^2+m_l^2-k_{jl}^2)/(2m_j m_l)$, and 
(in the spherical case) the integration extends over the spherical triangle 123 of the unit sphere,
see in figure~\ref{fig3pt1}b.

For the splitting we use the height of the basic tetrahedron, $m_0$, and obtain three tetrahedra,
as shown in figure~5a. 
One of them has the sides $m_1$, $m_2$, $m_0$, $K_{12}\equiv\sqrt{k_{12}^2}$, 
$K_{01}\equiv\sqrt{k_{01}^2}$ and $K_{02}\equiv\sqrt{k_{02}^2}$,
and the sides for the others can be obtained by permutation of the indices.
Here $k_{01}^2 = m_1^2 - m_0^2$, $k_{02}^2 = m_2^2 - m_0^2$, $k_{03}^2 = m_3^2 - m_0^2$, and
$m_0=m_1 m_2 m_3 \sqrt{D^{(3)}/\Lambda^{(3)}}$, where $\Lambda^{(3)} = {\textstyle{\frac{1}{4}}}
\left[ 2 k_{12}^2 k_{13}^2 + 2 k_{13}^2 k_{23}^2 + 2 k_{23}^2 k_{12}^2
- (k_{12}^2)^2 - (k_{13}^2)^2 - (k_{23}^2)^2 \right]$, and $D^{(3)}=\det\|c_{jl}\|$ is 
the Gram determinant, see in~\cite{DD-JMP,D-NIMA} for more details.
Each of the resulting integrals
can be associated with a specific three-point function, and we arrive at the following decomposition:
\begin{eqnarray}
J^{(3)}\!\left( n; 1,1,1\big| k_{23}^2, k_{13}^2, k_{12}^2; m_1, m_2, m_3 \!\right)
&\!\!\!\!=\!\!\!& 
\frac{m_1^2 m_2^2 m_3^2}{\Lambda^{(3)}}
\Biggl\{
\frac{F_1^{(3)}}{m_1^2}
J^{(3)}\!\left( n; 1,1,1\big| k_{23}^2, k_{03}^2, k_{02}^2; m_0, m_2, m_3 \!\right)
\hspace*{-2mm}
\nonumber \\ && 
+ \frac{F_2^{(3)}}{m_2^2}
J^{(3)}\!\left( n; 1,1,1\big| k_{03}^2, k_{13}^2, k_{01}^2; m_1, m_0, m_3 \right)
\nonumber \\ && 
+ \frac{F_3^{(3)}}{m_3^2}
J^{(3)}\!\left( n; 1,1,1\big| k_{02}^2, k_{01}^2, k_{12}^2; m_1, m_2, m_0 \right)
\Biggr\} \; ,
\label{J3_splitting1}
\end{eqnarray}
with
\begin{equation}
F_3^{(3)}
= \frac{1}{4m_1^2m_2^2}
\Bigl[
k_{12}^2 \left( k_{13}^2 \!+\! k_{23}^2 \!-\! k_{12}^2 
\!+\! m_1^2 \!+\! m_2^2 \!-\! 2 m_3^2\right)
- (m_1^2-m_2^2) \left( k_{13}^2 - k_{23}^2 \right)
\Bigr], 
\end{equation}
etc., so that 
$\sum_{i=1}^3 (F_i^{(3)}/m_i^2) = \Lambda^{(3)}/(m_1^2 m_2^2 m_3^2)$.

By dropping perpendiculars onto the sides $K_{12}\equiv\sqrt{k_{12}^2}$, etc., 
each of the resulting tetrahedra can be split into two, so that in
total we get six ``birectangular" tetrahedra, as shown in figure~5b.
In this way, we get the following relations for the integrals on the 
right-hand side of equation~(\ref{J3_splitting1}):
\begin{eqnarray}
&& \hspace*{-20mm}
J^{(3)}\left( n; 1,1,1 \big| k_{02}^2, k_{01}^2, 
k_{12}^2; m_1, m_2, m_0\right)
\nonumber \\
&=& 
\frac{1}{2 k_{12}^2}
\Bigl\{
(k_{12}^2 + m_1^2 - m_2^2)
J^{(3)}\left( n; 1,1,1 \big| k_{00'}^2, k_{01}^2, k_{10'}^2; m_1, m_{0'}, m_0\right)
\nonumber  \\
&& 
+ (k_{12}^2 - m_1^2 + m_2^2)
J^{(3)}\left( n; 1,1,1 \big| k_{02}^2, k_{00'}^2, k_{20'}^2; m_{0'}, m_2, m_0\right)
\Bigr\} \; ,
\end{eqnarray}
etc. The notation $0'$ is explained in figure~6;
in particular, $m_{0'}$ is the distance between the points $M$ and $0'$, whereas
$K_{10'}=\sqrt{k_{10'}^2}$ and $K_{20'}=\sqrt{k_{20'}^2}$
are the distances between the points $(1,0')$ and $(2,0')$, respectively,
so that $K_{10'}+K_{20'}=K_{12}$. Note that
$k_{10'}^2 = (k_{12}^2+m_1^2-m_2^2)^2/(4 k_{12}^2)$ and 
$k_{20'}^2 = (k_{12}^2-m_1^2+m_2^2)^2/(4 k_{12}^2)$,
similarly to the reduction of the two-point function.


\begin{wrapfigure}{rb}{0.5\textwidth}
\vspace*{-16mm} 
\begin{center}
\vspace*{-5mm}
\hspace*{-10mm}\includegraphics[width=22pc]{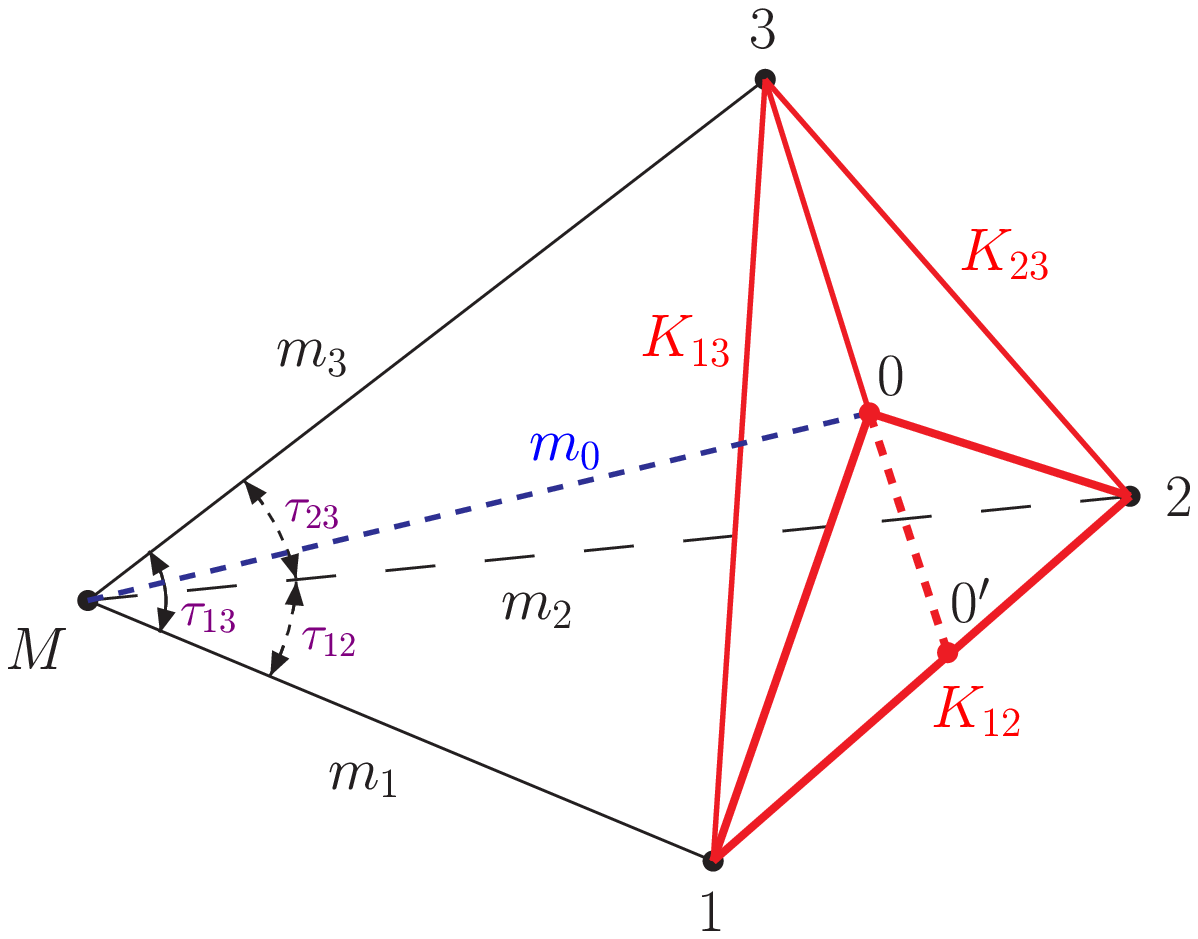}
\vspace*{-1mm}
\caption{\label{label}Last step of splitting the basic three-dimensional tetrahedron.}
\vspace*{-5mm}
\end{center}
  \label{fig3pt3}
\end{wrapfigure} 

Let us analyze the number of variables in the occurring three-point integrals. In 
$J^{(3)}\left( n; 1,1,1\big| k_{23}^2, k_{13}^2, k_{12}^2; m_1, m_2, m_3 \right)$ 
we have six independent variables: three masses and three momentum invariants 
(out of them we can construct five dimensionless variables).
In $J^{(3)}\left( n; 1,1,1 \big| k_{02}^2, k_{01}^2, 
k_{12}^2; m_1, m_2, m_0\right)$
we have two extra conditions on the variables, $k_{01}^2=m_1^2-m_0^2$ 
and $k_{02}^2=m_2^2-m_0^2$,
so that we get four independent variables (i.e., 
three dimensionless variables). 
For the integral
$J^{(3)}\left( n; 1,1,1 \big| k_{00'}^2, k_{01}^2, k_{10'}^2; m_1, m_{0'}, m_0\right)$
we have three relations, 
$k_{01}^2=m_1^2-m_0^2$, $k_{00'}^2 = k_{01}^2-k_{10'}^2$ 
and $k_{00'}^2 = m_{0'}^2-m_0^2$.
Therefore,
the result for the three-point function in arbitrary dimension 
should be expressible in terms of a combination of functions of
two dimensionless variables: indeed, we know that it can be presented
in terms of the Appell hypergeometric function $F_1$
(see, e.g., in~\cite{D-NIMA,Tarasov-NPBPS,FJT}).

This can also be seen in the integrands of 
the corresponding Feynman parametric integrals:
for the original three-point integral, the quadratic form is 
\begin{eqnarray}
&& \hspace*{-5mm}
J^{(3)}\left( n; 1,1,1\big| k_{23}^2, k_{13}^2, k_{12}^2; m_1, m_2, m_3 \right) 
\nonumber \\ 
&& \hspace*{35mm}
\Rightarrow
[ \alpha_1\alpha_2 k_{12}^2 + \alpha_1\alpha_3 k_{13}^2 + \alpha_2\alpha_3 k_{23}^2
- \alpha_1 m_1^2 - \alpha_2 m_2^2 - \alpha_3 m_3^2 ] \, ,
\end{eqnarray}
and for one of the resulting integrals after splitting,
remembering that $\alpha_1+\alpha_2+\alpha_3=1$, we get
\begin{eqnarray}
&& \hspace*{-5mm}
J^{(3)}\left( n; 1,1,1 \big| k_{00'}^2, k_{01}^2, 
k_{10'}^2; m_1, m_{0'}, m_0\right) 
\nonumber \\
&& \hspace*{35mm}
\Rightarrow
[ \alpha_1\alpha_2 k_{10'}^2 + \alpha_1\alpha_3 k_{01}^2 + \alpha_2\alpha_3 k_{00'}^2
- \alpha_1 m_1^2 - \alpha_2 m_{0'}^2 - \alpha_3 m_0^2 ]
\nonumber \\ 
&& \hspace*{35mm}
\Rightarrow
-[ \alpha_1^2 k_{10'}^2 + (\alpha_1 + \alpha_2)^2 k_{00'}^2 + m_0^2 ].
\end{eqnarray}
In this way, we obtain the following result in arbitrary dimension:
\noindent
\begin{eqnarray}
&& \hspace*{-10mm}
J^{(3)}\left( n; 1,1,1 \big| k_{00'}^2, k_{01}^2, 
k_{10'}^2; m_1, m_{0'}, m_0\right)
\nonumber \\
&=& 
-{\rm i} \pi^{n/2} \Gamma(3-n/2)
\int\limits_0^1 \!\!\! \int\limits_0^1 \!\!\! \int\limits_0^1
\frac{{\rm d}\alpha_1\; {\rm d}\alpha_2\; {\rm d}\alpha_3\; \delta(\alpha_1+\alpha_2+\alpha_3-1)}
{[ \alpha_1^2 k_{10'}^2 + (\alpha_1 + \alpha_2)^2 k_{00'}^2 + m_0^2 ]^{3-n/2}}
\nonumber \\
&=& -\frac{{\rm i}\pi^{n/2}\Gamma(2-n/2)}{2(m_0^2)^{2-n/2} k_{00'}^2}\;
\Biggl\{ \sqrt{\frac{k_{00'}^2}{k_{10'}^2}} \arctan\sqrt{\frac{k_{10'}^2}{k_{00'}^2}}
\nonumber \\
&& 
\hspace*{40mm}
- \left(\frac{m_0^2}{m_{0'}^2}\right)^{2-n/2}
F_1\left( 1/2, 1, 2-n/2; 3/2 \Bigl| -\frac{k_{10'}^2}{k_{00'}^2}, -\frac{k_{10'}^2}{m_{0'}^2} \right)
\Biggl\} ,
\label{J3_split_F1}
\end{eqnarray}
where $F_1$ is Appell hypergeometric function of two variables,
\[
F_1\left( a, b_1, b_2; c | x, y\right) = \sum\limits_{j_1, j_2} 
\frac{(a)_{j_1+j_2}\; (b_1)_{j_1}\; (b_2)_{j_2}}{(c)_{j_1+j_2}}\;
\frac{x^{j_1}\; y^{j_2}}{j_1!\; j_2!} \; .
\]
Similar results for other five contributions can be obtained by permutation.
Using known transformation formulae for $F_1$ we can see that the obtained 
expression~(\ref{J3_split_F1}) is equivalent to the result presented in~\cite{D-NIMA}.

\section{Four-point function}

For the four-point function (see figure~\ref{fig234pt}c), 
there are six external momentum invariants. 
Out of them, $k_{12}^2$, $k_{23}^2$, $k_{34}^2$ and $k_{14}^2$ are the squared
momenta of the external legs, whilst $k_{13}^2$ and $k_{24}^2$ correspond
to the Mandelstam variables $s$ and $t$. The sides of the corresponding basic 
four-dimensional simplex are $m_1$, $m_2$, $m_3$, $m_4$, and six additional
sides $K_{jl}\equiv\sqrt{k_{jl}^2}$,
as shown in figure~\ref{fig4pt1}a. The six angles $\tau_{jl}$
between the corresponding sides $m_j$ and $m_l$
are defined through $\cos\tau_{jl}\equiv c_{jl}=(m_j^2+m_l^2-k_{jl}^2)/(2m_j m_l)$, and 
(in the spherical case) the integration extends over the spherical tetrahedron 1234 
of the unit hypersphere, as shown in figure~\ref{fig4pt1}b (for the hyperbolic case
one can use analytic continuation).

\begin{figure}[h]
\vspace*{10mm}
\begin{center}
\includegraphics[width=30pc]{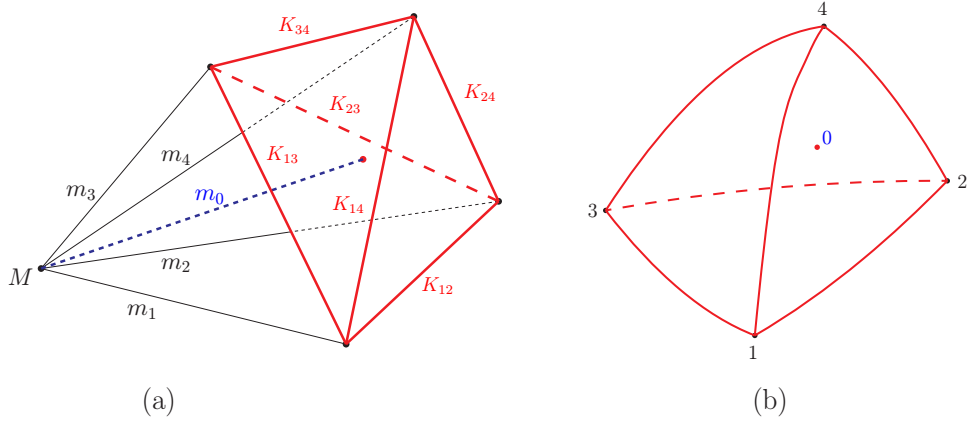}
\end{center}
  \label{fig4pt1}
\vspace*{-15mm}
\caption{\label{label}Four-point case: (a) the basic simplex 
and (b) the spherical tetrahedron.}
\label{fig4pt1} 
\end{figure}

\begin{figure}[h]
\vspace*{10mm}
\includegraphics[width=30pc]{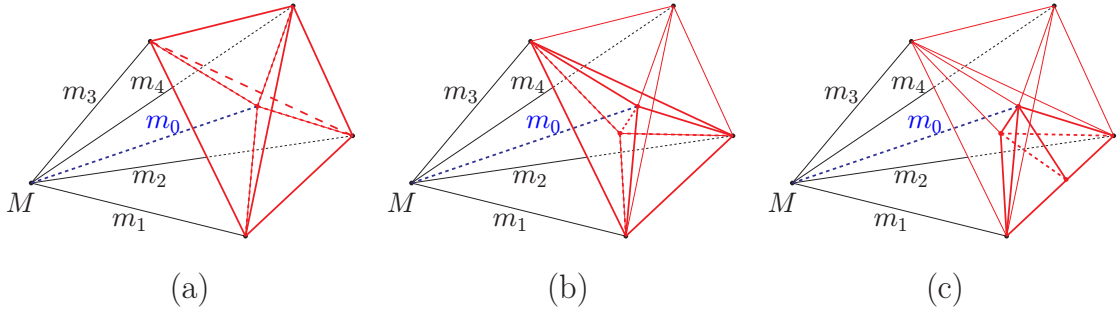}
\vspace*{-3mm}
\caption{\label{label}Four-point case: steps of splitting the basic four-dimensional simplex.}
\label{fig4pt2}
\end{figure}


For splitting we use the height of the basic simplex, $m_0$, and obtain four simplices,
as shown in figures~\ref{fig4pt2} and \ref{fig4pt2}a. 
One of them has the sides $m_1$, $m_2$, $m_3$, $m_0$, $K_{12}\equiv\sqrt{k_{12}^2}$, 
$K_{13}\equiv\sqrt{k_{13}^2}$, $K_{23}\equiv\sqrt{k_{23}^2}$, 
$K_{01}\equiv\sqrt{k_{01}^2}$, $K_{02}\equiv\sqrt{k_{02}^2}$ and $K_{03}\equiv\sqrt{k_{03}^2}$,
and the sides of the others can be obtained by permutation of the indices.
As before, $k_{0i}^2 = m_i^2 - m_0^2$ ($i=1,2,3,4$), whereas
$m_0=m_1 m_2 m_3 m_4\sqrt{D^{(4)}/\Lambda^{(4)}}$, where $D^{(4)}=\det\|c_{jl}\|$ and 
$\Lambda^{(4)} = \det\|(k_{j4}\cdot k_{l4})\|$, 
see in~\cite{DD-JMP} for more details.
Each of the four resulting integrals
can be associated with a certain four-point function.
At the next step, in each of the four tetrahedra (drawn in red) we drop the perpendiculars 
onto the triangle sides, as shown in figure~\ref{fig4pt2}b, splitting each of them into three, 
and then dividing each of the resulting
tetrahedra into two, by dropping perpendiculars onto the $\sqrt{k_{jl}^2}$ sides, 
as shown in figure~\ref{fig4pt2}c.
As a result of this splitting, we get $4\cdot3\cdot2=24$ simplices.

\begin{wrapfigure}{rb}{0.5\textwidth}
\vspace*{-5mm} 
\begin{center}
\vspace*{-5mm}
\hspace*{-10mm}\includegraphics[width=20pc]{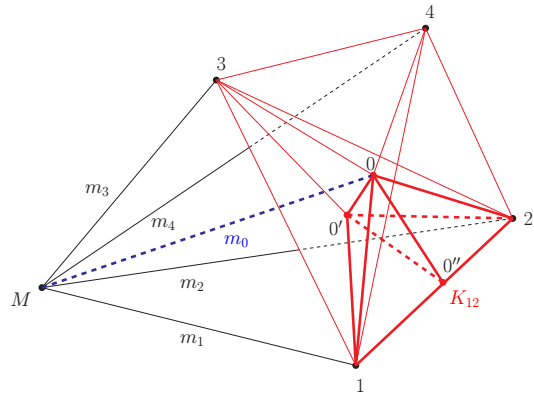}
\vspace*{-1mm}
\caption{\label{label}Last step of splitting the basic four-dimensional simplex.}
\vspace*{-9mm}
\end{center}
\label{fig4pt4a}
\end{wrapfigure} 

Let us look at the number of variables. In the integral 
$J^{(4)}\left( n; 1,1,1,1\big| \{k_{jl}^2\}; \{m_i\} \right)$ 
we have ten independent variables: four masses and six momentum invariants 
(out of them we can construct nine dimensionless variables).
After the first step (figure~\ref{fig4pt2}a)
we have three conditions on the variables, $k_{01}^2=m_1^2-m_0^2$, $k_{02}^2=m_2^2-m_0^2$ 
and $k_{03}^2=m_3^2-m_0^2$,
so that we get seven independent variables (i.e., 
six dimensionless variables). 
After the second step (figure~\ref{fig4pt2}b), we get two extra conditions due to the right triangles,
and after the third step (figure~\ref{fig4pt2}c) we get one more condition.
As a result, for each of the 24 resulting four-point functions we have six relations,
so that we end up with four independent variables
(i.e., three dimensionless variables). 
Therefore, the result for the four-point function in arbitrary dimension 
should be expressible in terms of a combination of functions of
three dimensionless variables, such as, e.g., Lauricella functions and 
their generalizations (see, e.g., in~\cite{FJT,BKM}).

%

This can also be seen in the integrands of 
the corresponding Feynman parametric integrals:
for the original four-point integral, the quadratic form is 
\begin{eqnarray}
&& \hspace*{-5mm}
J^{(4)}\left( n; 1,1,1,1 \big| \{ k_{12}^2, k_{23}^2, k_{34}^2, k_{14}^2, k_{13}^2, k_{24}^2\}; 
\{m_1, m_2, m_3, m_4\} \right)
\nonumber \\
&& \hspace*{30mm}
\Rightarrow 
[\alpha_1 \alpha_2 k_{12}^2 + \alpha_1 \alpha_3 k_{13}^2 + \alpha_1 \alpha_4 k_{14}^2 
+ \alpha_2 \alpha_3 k_{23}^2 + \alpha_2 \alpha_4 k_{24}^2 + \alpha_3 \alpha_4 k_{34}^2 
\nonumber \\
&& \hspace*{40mm}
- \alpha_1 m_1^2 - \alpha_2 m_2^2 - \alpha_3 m_3^2 - \alpha_4 m_4^2 ] \, ,
\end{eqnarray}
whereas after the last step of splitting,
remembering that $\alpha_1+\alpha_2+\alpha_3+\alpha_4=1$, we get
\begin{eqnarray}
&& \hspace*{-5mm}
J^{(4)}\left( n; 1,1,1,1 \big| 
\{ k_{10''}^2, k_{0'0''}^2, k_{00'}^2, k_{01}^2, k_{10'}^2, k_{00''}^2\}; 
\{m_1, m_{0''}, m_{0'}, m_0\} \right)
\nonumber \\
&& \hspace*{30mm}
\Rightarrow
[\alpha_1 \alpha_2 k_{10''}^2 + \alpha_1 \alpha_3 k_{10'}^2 + \alpha_1 \alpha_4 k_{01}^2 
+ \alpha_2 \alpha_3 k_{0'0''}^2 + \alpha_2 \alpha_4 k_{00''}^2 + \alpha_3 \alpha_4 k_{00'}^2 
\nonumber \\
&& \hspace*{40mm}
- \alpha_1 m_1^2 - \alpha_2 m_{0''}^2 - \alpha_3 m_{0'}^2 - \alpha_4 m_0^2 ]
\nonumber \\
&& \hspace*{30mm}
\Rightarrow 
- [\alpha_1^2 k_{10''}^2 + (\alpha_1+\alpha_2)^2 k_{0'0''}^2 
+ (\alpha_1+\alpha_2+\alpha_3)^2 k_{00'}^2 + m_0^2 ] \, .
\end{eqnarray}
The notations $0'$ and $0''$ are explained in figure~9;
in particular, $m_{0'}$ is the distance between $M$ and $0'$,
$m_{0''}$ is the distance between $M$ and $0''$, whereas
$K_{00'}=\sqrt{k_{00'}^2}$, $K_{0'0''}=\sqrt{k_{0'0''}^2}$ 
and $K_{10''}=\sqrt{k_{10''}^2}$
are the distances between the corresponding points $(0,0')$, $(0',0'')$ and $(1,0'')$, 
respectively.
In this way, we obtain the following result in arbitrary dimension:
\begin{eqnarray}
&& \hspace*{-15mm}
J^{(4)}\left( n; 1,1,1,1 \big| 
\{ k_{10''}^2, k_{0'0''}^2, k_{00'}^2, k_{01}^2, k_{10'}^2, k_{00''}^2\}; 
\{m_1, m_{0''}, m_{0'}, m_0\} \right)
\nonumber \\ &=& \!\!\! 
{\rm i} \pi^{n/2} \Gamma(4\!-\!n/2)
\int\limits_0^1 \!\!\! \int\limits_0^1 \!\!\! \int\limits_0^1 \!\!\! \int\limits_0^1
\frac{{\rm d}\alpha_1\; {\rm d}\alpha_2\; {\rm d}\alpha_3\; {\rm d}\alpha_4\; 
\delta(\alpha_1+\alpha_2+\alpha_3+\alpha_4-1)}
{[\alpha_1^2 k_{10''}^2 + (\alpha_1\!+\!\alpha_2)^2 k_{0'0''}^2 
+ (\alpha_1\!+\!\alpha_2\!+\!\alpha_3)^2 k_{00'}^2 + m_0^2 ]^{4-n/2}}
\nonumber \\ &=& \!\!\! 
\frac{{\rm i} \pi^{n/2} \Gamma(3\!-\!n/2)}{2 k_{0'0''}^2 (m_0^2)^{3-n/2}}
\Biggl\{ \sqrt{\frac{k_{0'0''}^2}{k_{10''}^2}}\arctan\sqrt{\frac{k_{10''}^2}{k_{0'0''}^2}}\;\;
{}_2F_1\left(
\begin{array}{c} 1/2, 3-n/2\\ 3/2 \end{array} \Bigl| -\frac{k_{00'}^2}{m_0^2} 
\right)
\\ && \hspace*{-5mm}
- \left( \!\frac{m_0^2}{m_{0'}^2}\! \right)^{\!\!2-n/2}\!\!\!
F_N\!\!\left( \!1, 1, 3\!-\!n/2, 1/2, (n\!-\!3)/2, 1/2; 3/2, 3/2, 3/2 \Bigl| 
-\frac{k_{10''}^2}{k_{0'0''}^2}, -\frac{k_{00'}^2}{m_0^2}, -\frac{k_{10'}^2}{m_{0'}^2}
\right)\!\!\!
\Biggl\},
\end{eqnarray}
where $F_N$ is one of the Lauricella-Saran functions~\cite{Saran2},
\[
F_N(a_1, a_2, a_3, b_1, b_2, b_1; c_1,c_2, c_2 | x, y, z)
= \sum\limits_{j_1, j_2, j_3} \!\!\! \frac{(a_1)_{j_1} (a_2)_{j_2} (a_3)_{j_3} (b_1)_{j_1+j_3} (b_2)_{j_2}}
{(c_1)_{j_1} (c_2)_{j_2+j_3}} \; \frac{x^{j_1} y^{j_2} z^{j_3}}{j_1! j_2! j_3!} \, .
\]

\section{General remarks and conclusions}

The geometrical approach allows us to
relate the one-loop $N$-point Feynman diagrams to certain 
(hyper)volume integrals in non-Euclidean geometry. 
Geometrical splitting provides a straightforward way of reducing
general integrals to those with lesser number of independent
variables. Furthermore, in this way we can predict the set and 
the number of these variables
in the resulting integrals. As shown in~\cite{D-JPCS}, for an $N$-point
diagram (depending, in the general off-shell case, on
$\tfrac{1}{2}(N-1)(N+2)$ dimensionless variables), after splitting in $N!$ pieces
we will get a combination of $N!$ integrals, each of them depending only on
$N-1$ variables. For example, in the four-point case we will get functions of
three variables, rather than nine.  

%

Geometrically, we can calculate the resulting integrals in the framework of 
non-Euclidean geometry or, alternatively, represent them again in terms of Feynman 
parameters. Since some of the variables are connected, the quadratic forms
in the integrands of the resulting parametric integrals can be simplified. 
In this way, for $N=2, 3, 4$ we get the following quadratic forms:
\begin{eqnarray*}
&& \hspace*{-10mm}
J^{(2)}\left( n; 1,1 \big| k_{01}^2; m_1, m_0\right) 
\Rightarrow - [ \alpha_1^2 k_{01}^2 + m_0^2 ] \, ,
\\
&& \hspace*{-10mm}
J^{(3)}\left( n; 1,1,1 \big| k_{00'}^2, k_{01}^2, 
k_{10'}^2; m_1, m_{0'}, m_0\right)
\Rightarrow -[ \alpha_1^2 k_{10'}^2 + (\alpha_1 + \alpha_2)^2 k_{00'}^2 + m_0^2 ] \, ,
\\
&& \hspace*{-10mm}
J^{(4)}\left( n; 1,1,1,1 \big| 
\{ k_{10''}^2, k_{0'0''}^2, k_{00'}^2, k_{01}^2, k_{10'}^2, k_{00''}^2\}; 
\{m_1, m_{0''}, m_{0'}, m_0\} \right)
\\
&& \hspace*{20mm}
\Rightarrow - [\alpha_1^2 k_{10''}^2 + (\alpha_1+\alpha_2)^2 k_{0'0''}^2  
+ (\alpha_1+\alpha_2+\alpha_3)^2 k_{00'}^2 + m_0^2 ] \, .
\end{eqnarray*}
Evaluating these integrals for an arbitrary dimension $n$ we can get explicit expressions 
in terms of (generalized) hypergeometric functions: ${}_2F_1$ for $N=2$, $F_1$ for $N=3$,
and $F_N$ for $N=4$.
For $N>4$, in the quadratic forms we should also expect sums of squares of partial sums of $\alpha$'s
(the coefficient of $m_0^2$ can be understood as the square 
of the sum of all $\alpha$'s, equal to one).

\section*{Acknowledgements}
I am thankful to R~Delbourgo and M~Yu~Kalmykov with whom I started to work on this subject. I am 
grateful to the organizers of ACAT-2017 for their support and hospitality.

\end{document}